# NN Potentials from Nonlinear Quantum Field Theory

Lutz Jäde and Heinrich Viktor v. Geramb

*Theoretische Kernphysik, Universität Hamburg*
*Luruper Chaussee 149, D-22761 Hamburg*
e-mail jae@i04kthd.desy.de

October 1995

## Abstract

We investigate self–interacting scalar, pseudoscalar and vector meson fields and their influence on NN interactions. Due to the self–interaction one has to solve nonlinear field equations which allow solitary wave solutions. A propability amplitude for the propagation of these solutions is calculated which can be applied in a One–Boson–Exchange Potential (OBEP). Using proper normalization as an alternative to the usual renormalization scheme one obtains amplitudes with the same momentum dependence as the conventional Feynman–propagators multiplied with the form factors of the Bonn–B OBEP. From this the usage of form factors might be replaced by the application of solitary wave exchange. In addition the parameters entering the different meson propagators show a connection in form of a simple scaling law which relates all parameters to the pion self–interaction coupling constant.

# 1 The Meson Self–Interaction

The fact that the Lagrangian for meson fields contains nonlinear powers of the field operators is well established in nuclear physics [1]. The detailed structure of these terms might be determined by certain symmetry conditions e.g. isospin and chiral invariance. Rather than to derive the form of the Lagrangian our goal is to test the influence of the nonlinearities on observable data. From this we focus on the simplest possible form for the field equations which follow from the Lagrangians for pseudoscalar and scalar mesons. We regard mesons which carry out a self–interaction of polynomial form leading to a nonlinear field equation for the meson field operators:

$$\partial_\mu \partial^\mu \Phi(x,k) + m^2 \Phi(x,k) + \lambda \Phi^{2p+1}(x,k) = 0, \tag{1}$$

where $\lambda$ is the coupling constant of the self–interaction and $k$ is the momentum of the field. The case $p = 1/2$ will lead to scalar and $p = 1$ to pseudoscalar self–interaction. Special solutions of this nonlinear equation are the *solitary wave solutions* which are wave–like and depend on the free solutions $\varphi(x,k)$ which solve the equation for $\lambda = 0$:

$$\partial_\mu \partial^\mu \varphi(x,k) + m^2 \varphi(x,k) = 0. \tag{2}$$

This is the well–known Klein–Gordon equation. The free wave–like solutions in a finite volume $V$ are:

$$\varphi(x,k,\pm) = \frac{1}{\sqrt{2D_k \omega_k V}} \, a(k,\pm) \, e^{\mp ikx}, \tag{3}$$

where the operators $a(k,\pm)$ are the annihilation operators for positive and negative energy $\omega_k$. The arbitrary constant $D_k$ can depend on the energy $\omega_k$ and the coupling constant $\lambda$ with the constraint of Lorentz–invariance.

Making the ansatz $\Phi = \Phi(\varphi)$ one can solve (1) by direct integration [2]:

$$\Phi = \varphi \left[ 1 - \frac{\lambda \varphi^{2p}}{4(p+1)m^2} \right]^{-\frac{1}{p}}. \tag{4}$$

These solutions are expressed as a power series

$$\Phi(x,k,\pm) = \sum_{n=0}^{\infty} C_n^{1/2p}(1) \left( \frac{\lambda}{4(p+1)m^2} \right)^n \varphi^{2pn+1}(x,k,\pm) \tag{5}$$

where $C_n^a(x)$ are Gegenbauer Polynomials satisfying:

$$C_n^{1/2}(1) = P_n(1) = 1 \qquad C_n^1(1) = n + 1.$$

The solutions (3) and therefore the operator $\Phi(\varphi)$ are quantized by the commutator relation [3]:

$$[a(k), a^\dagger(k')] = \delta_{kk'}. \tag{6}$$

Using this relation one can construct a complete set of orthonormal eigenvectors to the number operator:

$$N_k = a^\dagger(k) a(k)$$

which read:

$$|n_k\rangle = \frac{1}{\sqrt{n!}} a^{\dagger n}(k) |0\rangle.$$

The commutator (6) leads to:

$$\langle 0|a^n(k)a^{\dagger m}(k)|0\rangle = n! \cdot \delta_{nm}. \tag{7}$$

Now one can calculate the matrix elements of the operator (5). If one asks for the probability to find $m$–particles with momentum $k^\mu$ when $\Phi^\dagger(x,k)$ has operated on the $n$–particle state at $x$ it follows:

$$\langle m_k|\Phi^\dagger(x,k)|n_k\rangle =$$

$$\sum_l \frac{C_l^{\frac{1}{2p}}(1)}{\sqrt{m!n!}} \left(\frac{1}{\sqrt{2\omega D_k V}}\right)^{2pl+1} \left(\frac{\lambda}{4(p+1)m^2}\right)^l \langle 0|a^m(k)a^{\dagger 2pl+1+n}(k)|0\rangle \, e^{ikx(2pl+1)}. \tag{8}$$

Due to (7) the only nonvanishing term includes:

$$l = \frac{m-n-1}{2p}.$$

Since $l$ has to be an integer this relation can only be fulfilled when the difference $m-n$ is an integer greater than zero which has to be even for $p=1/2$ and odd for $p=1$, otherwise the matrix element vanishes. The nonzero elements are:

$$\langle m_k|\Phi^\dagger(x,k)|n_k\rangle_{m-n=2pl+1} = \sqrt{\frac{m!}{n!}}\, C_l^{\frac{1}{2p}}(1) \left(\frac{1}{\sqrt{2\omega D_k V}}\right)^{m-n} \left(\frac{\lambda}{4(p+1)m^2}\right)^{\frac{m-n-1}{2p}} e^{ikx(m-n)}. \tag{9}$$

Analogous one can ask for the probability to find a state which contains $m$ particles after $\Phi(y,k)$ has act on the state $|n_k\rangle$. Taking the adjoint of (9) yields:

$$\langle n_k|\Phi(y,k)|m_k\rangle_{m-n=2pl+1} = \sqrt{\frac{m!}{n!}}\, C_l^{\frac{1}{2p}}(1) \left(\frac{1}{\sqrt{2\omega D_k V}}\right)^{m-n} \left(\frac{\lambda}{4(p+1)m^2}\right)^{\frac{m-n-1}{2p}} e^{-iky(m-n)}. \tag{10}$$

The propagator for particles is defined as the probability for a positive energy particle to be created from the vacuum at $x$ and propagate forward in time to $y$ plus the probability for a negative energy particle to be created at $y$ and propagate backwards in time to $x$ where it is annihilated into the vacuum. The latter might be described by a positive energy antiparticle which propagates forward in time from $y$ to $x$. Using the amplitudes (9) and (10) a probability amplitude $iP(y-x)$ for the propagation of the interacting system analogous to the particle propagator can be calculated. Since one wants to get the probability for the propagation of a solitary wave one has to use the operator $\Phi(x,k)$ to annihilate and create the propagating quanta. This operator contains a series of $(2pn+1)$–particle states and so there exist propabilities to find $(2pn+1)$ particles between the creation and the annihilation [fig. 1]. The intermediate state is not measurable and so a sum over all possible intermediate states has to be performed. The particles are undistinguishable and therefore the propability to find a $(2pn+1)$–particle state has to be weighted by a combinatorial factor [4]. Due to (6) the momentum $k$ has to be the same for $\Phi$ and $\Phi^\dagger$ and a sum runs over all momenta $k$ has to be performed. Summation over all $n$ yields:

$$iP(y-x) = \sum_{n=0}^\infty iP_n(y-x)$$

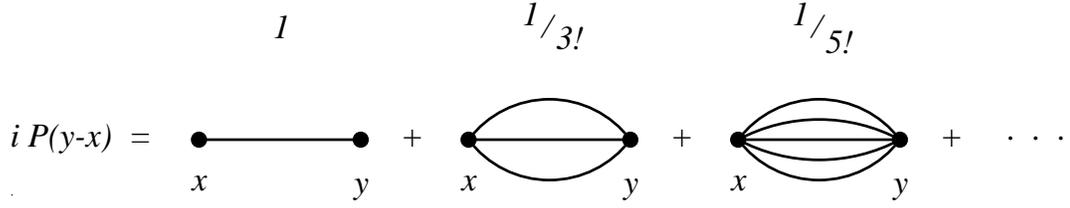

Figure 1: The $\lambda\Phi^3$ propagator as a sum over different probability amplitudes

Where:

$$iP_n(y-x) = \frac{1}{(2pn+1)!}\sum_k \Big[\langle 0|\Phi(y,k)|2pn+1,k\rangle\langle 2pn+1,k|\Phi^\dagger(x,k)|0\rangle\theta(y_0-x_0)$$
$$+ \langle 0|\Phi(x,k)|2pn+1,k\rangle\langle 2pn+1,k|\Phi^\dagger(y,k)|0\rangle\theta(x_0-y_0)\Big] \quad (11)$$

Using (9) and (10) and taking the limit of large $V$ the sum over $k$ goes into an integral. Together with the integral representation of the $\theta$–functions one gets:

$$iP(y-x) = \frac{i}{(2\pi)^4}\int d^4k \; P(k^2,m)\, e^{-ik(y-x)}.. \quad (12)$$

Where the momentum space amplitude reads:

$$iP(k^2,m) = \sum_{n=0}^{\infty}\left[C_n^{1/2p}(1)\right]^2 \frac{(2pn+1)^{2pn-2}(m^p\alpha)^{2n}}{D_k^{2pn+1}(\vec{k}^{\,2}+M_n^2)^{pn}}\, i\Delta_F(k^2,M_n) \quad (13)$$

with the Feynman propagator:

$$i\Delta_F(k^2,M_n) = \frac{i}{k_\mu k^\mu - M_n^2} \quad (14)$$

and a mass–spectrum:

$$M_n = (2pn+1)m.$$

The dimensionless coupling constant $\alpha$ is defined by:

$$\alpha = \frac{\lambda}{4(p+1)m^2(2mV)^p}.$$

This amplitude shall be referred to as 'modified solitary wave propagator' and it can now be used in meson field theories.

## 1.1 Proper Normalization

Since the factor $D_k$ is one of the two integration constants of the second order differential equation (1) and can depend on $\omega_k$ and $\vec{k}$ it has to be fixed by physical boundary conditions. These conditions are [2]:

1. All amplitudes must be Lorentz–invariant.

2. $D_k$ must be dimensionless.

3. All self–scattering diagramms must be finite.

4. The fields have to vanish for $\lambda \to 0$

Whereas the first and second condition demand that $m$, $\lambda$ (dimension $[m^{-2p+2}]$), $\omega_k$ and $V$ (dimension $[m^{-3}]$) just can appear in the combination

$$\left(\frac{m^2}{\lambda}\right)^{\frac{2}{p}} \omega_k \cdot V = \left(\frac{m^2}{\lambda}\right)^{\frac{2}{p}} \sqrt{\vec{k}^{\,2} + m^2} \cdot V$$

which is a dimensionless Lorentz–scalar the third condition is similar to the renormalization scheme of the linear model. The last condition makes the nonlinear aspects of the fields depend completely on the interaction. No interaction means no fields [2].

To find a constraint on the $\vec{k}$–dependence of $D_k$ one has to look on the worst case of a self–scattering diagramm. For the $\lambda \Phi^3$ current this is the first correction to the two–point function $iP(k, m)$ which reads:

$$iP'(k, m) \sim \left(iP(k, m)\right)^2 \int d^4 p_1 d^4 p_2 \; iP(p_1, m) iP(p_2, m) iP(k - p_1 - p_2, m) \qquad (15)$$

Inserting the series (21) it is the term for $n = 0$ which shows the worst behaviour and grows for large momenta like:

$$d^4 p_1 d^4 p_2 \frac{1}{D_k(p_1) D_k(p_2) D_k(k - p_1 - p_2) \cdot p_1^2 \cdot p_2^2 \cdot (k - p_1 - p_2)^2}.$$

Since each integrand can behave like a constant in some regions of momentum space namely where the momentum vectors satisfy:

$$|\vec{k} - \vec{p}_1| \sim |\vec{p}_2|.$$

the factor $D_k$ has to be chosen so that each integral over $p_i$ remains finite. This leads to:

$$D_k \sim (\omega_k V)^s \qquad s \geq 2.$$

Simultanously $D_k$ has to be dimensionless and together with the constraint of $\Phi \to 0$ for $\lambda \to 0$ one gets the simplest form for $D_k$ to fulfill all three conditions [2]:

$$D_k = 1 + \left(\frac{m^2}{\lambda}\right)^{\frac{2}{p}} (\omega_k V)^2 = 1 + \frac{1}{\left(4(p+1)2^p \alpha\right)^{2/p}} \cdot \frac{\vec{k}^{\,2} + m^2}{m^2}. \qquad (16)$$

With this choice the modified solitary wave propagator (13) is now totally determined.

## 2 The Propagator in NN Scattering

In the conventional Boson–Exchange models one uses linear meson fields which are solutions of the Klein–Gordon equation. In our approach we want to apply the model of self–interacting mesons to obtain a solitary wave exchange potential (SWEP). The only difference to the common OBEPs lies in the meson dynamics and will manifest itself in the propagator. Therefore it is in the first step sufficient just to compare the mesonic part of the linear and the nonlinear model instead of calculating phase shifts. As a guiding line we will use the propagators applied in the Bonn–B potential [5, 6] which yields a good quantitative description of NN scattering data.

Table 1: Bonn–B Meson Parameters

|  | $\pi$ | $\eta$ | $\rho$ | $\omega$ | $\sigma_1$ | $\sigma_0$ | $\delta$ |
|---|---|---|---|---|---|---|---|
| $S^P$ | $0^-$ | $0^-$ | $1^-$ | $1^-$ | $0^+$ | $0^+$ | $0^+$ |
| $m_\beta$ (MeV) | 138.03 | 548.8 | 769 | 782.6 | 550 | 720 | 983 |
| $\Lambda_\beta$ (GeV) | 1.7 | 1.5 | 1.85 | 1.85 | 1.9 | 1.9 | 2.0 |
| $n_\beta$ (MeV) | 1 | 1 | 2 | 2 | 1 | 1 | 1 |

## 2.1 Linear Mesons – The Bonn–B Potential

In the conventional approach the mesons are solutions of a linear Klein–Gordon equation with physical mass $m$:

$$\partial_\mu \partial^\mu \varphi(x,k) + m^2 \varphi(x,k) = 0. \qquad (17)$$

This assumption, taking the self–interaction to affect the mass of the meson only, leads to a free propagator for the meson fields:

$$i\Delta_F(k^2, m) = \frac{i}{k_\mu k^\mu - m^2} \qquad (18)$$

In the Bonn–OBEP this propagator is used at $k_0 = 0$ [5]:

$$i\Delta_F(\vec{k}^2, m) = -\frac{i}{\vec{k}^2 + m^2}$$

In addition a form factor is applied to each meson–nucleon vertex which contains the cut–off masses of the mesons ($\beta = \pi, \sigma, \eta, \rho, \omega, \delta$):

$$i\Delta_F(k, m_\beta) \rightarrow \left[F_\beta(k)\right]^2 \cdot i\Delta_F(k, m_\beta) \qquad \text{with:} \qquad F_\beta = \left(\frac{\Lambda_\beta^2 - m_\beta^2}{\Lambda_\beta^2 + \vec{k}^2}\right)^{n_\beta} \qquad (19)$$

The form factor decreases with increasing energy and makes the momentum space potentials fall down to zero for increasing nucleon momenta. The cut–off masses $\Lambda_\beta$ and $n_\beta$ are adjusted to fit the observable data (see table [1]).

## 2.2 Nonlinear Mesons – Solitary Waves

Instead of treating the meson fields as linear fields with renormalized mass we assume a persistent self–interaction of cubic form for pseudoscalar mesons ($\pi, \eta$):

$$\partial_\mu \partial^\mu \Phi(x,k) + m_{ps}^2 \Phi(x,k) + \lambda_{ps} \Phi^3(x,k) = 0. \qquad (20)$$

Setting $p = 1$ and $k_0 = 0$ formula (13) yields the modified solitary wave propagator

$$iP_{ps}(\vec{k}^2, m_{ps}) = -i \sum_{n=0}^{\infty} \frac{(2n+1)^{2n-2}(m_{ps}\alpha_{ps})^{2n}}{D_{k,ps}^{2n+1}} \cdot \frac{1}{\left[\vec{k}^2 + (2n+1)^2 m_{ps}^2\right]^{n+1}}. \tag{21}$$

with:

$$\alpha_{ps} = \frac{\lambda_{ps}}{16 m_{ps}^3 V}$$

and:

$$D_{k,ps} = 1 + \frac{1}{256} \cdot \frac{\vec{k}^2 + m_{ps}^2}{(m_{ps}\alpha_{ps})^2}. \tag{22}$$

For the vector mesons $\rho$ and $\omega$ which have negative intrinsic parity we assume a field equation of the form:

$$\partial_\mu \partial^\mu \Phi_v^\nu(x,k) + m_v^2 \Phi_v^\nu(x,k) + \lambda_v \Phi_{v\mu} \Phi_v^\mu \Phi_v^\nu(x,k) = 0. \tag{23}$$

Introducing the polarisation vector [3]:

$$\Phi_v^\nu(x,k) = \epsilon^\nu(k) \cdot \Phi_v(x,k) \qquad \text{with:} \quad \epsilon_\mu(k)\epsilon^\mu(k) = 1$$

the fields $\Phi_v(x,k)$ satisfy the equation (20) and according to Machleidt et al. we therefore split the propagator for the vector meson fields into a Minkowski tensor and a propagator function which has the same form as the pseudoscalar propagator (21):

$$iP_v^{\mu\nu}(k, m_v) = \left(-g^{\mu\nu} + \frac{k^\mu k^\nu}{m_v^2}\right) \cdot iP_{ps}(k, m_v). \tag{24}$$

For the scalar mesons $(\sigma_1, \sigma_0, \delta)$ we assume a quadratic self interaction. Setting $p = 1/2$ one gets:

$$\partial_\mu \partial^\mu \Phi(x,k) + m_s^2 \Phi(x,k) + \lambda_s \Phi^2(x,k) = 0 \tag{25}$$

leading to a propagator which reads:

$$iP_s(\vec{k}^2, m_s) = -i \sum_{n=0}^{\infty} \frac{(n+1)^{n-1}(\sqrt{m_s}\alpha_s)^{2n}}{D_{k,s}^{n+1}} \cdot \frac{1}{\left[\vec{k}^2 + (n+1)^2 m_s^2\right]^{n/2+1}} \tag{26}$$

with:

$$\alpha_s = \frac{\lambda_s}{6 m_s^2 \sqrt{2m_s V}}$$

and:

$$D_{k,s} = 1 + \frac{1}{5184} \cdot \frac{\vec{k}^2 + m_s^2}{(\sqrt{m_s}\alpha_s)^4} \tag{27}$$

These propagators are now compared to the Bonn–B propagators multiplied with the form factors. The identification of the modified solitary wave propagators with the Bonn–OBEP Feynman propagators *and* the meson–nucleon form factors reflects the idea that the self–interaction with proper normalization $D_k$ of the free meson fields is sufficient to describe the meson dynamics so that no empirical form factors are needed.

Table 2: PSWEP self–interaction coupling constants (masses in MeV)

|  | $\pi$ | $\eta$ | $\rho$ | $\omega$ | $\sigma_1$ | $\sigma_0$ | $\delta$ |
|---|---|---|---|---|---|---|---|
| $S^P$ | $0^-$ | $0^-$ | $1^-$ | $1^-$ | $0^+$ | $0^+$ | $0^+$ |
| $m_\beta$ | 138.03 | 548.8 | 769 | 782.6 | 550 | 720 | 983 |
| $\alpha_\beta$ | 0.36 | 0.1079 | 0.0580 | 0.0560 | 0.1755 | 0.151 | 0.1269 |

# 3 Results

To test if and how the effects of the self–interaction can be seen in nucleon–nucleon scattering data we compare the proper normalized modified solitary wave propagator with the Feynman propagators multiplied with the form factors (19) using the Bonn–B parameter set of table [1]. The treatment of the meson dynamics is the only difference between Bonn–OBEP and the solitary wave exchange potential SWEP. In both potentials the functions describing the meson propagation enter the potential in the same way.

In figure [2], [3] and [4] we plot the Bonn–B propagator multiplied with $i$ times the form factor compared with $i$ times the propagator (21) and (26) respectively. The coupling constants entering the solitary wave propagators are given in table [2]. Thus we compare:

$$iP(\vec{k}^{\,2}, m_\beta, \alpha_\beta) \quad \longleftrightarrow \quad \left(\frac{\Lambda_\beta^2 - m_\beta^2}{\Lambda_\beta^2 + \vec{k}^{\,2}}\right)^{2n_\beta} i\Delta_F(\vec{k}^{\,2}, m_\beta).$$

As a result we obtain qualitative agreement for all type of mesons. It is important in this context that the propagator follows from the field equation independent from its application in nucleon–nucleon scattering. Therefore the only parameter which enters the model is the coupling constant of the self interaction. It is obvious that the coupling constant decreases with increasing mass. This behaviour can be expressed in an empirical scaling law:

$$\alpha(m) \sim \text{const} \cdot m^{-\kappa}$$

Looking at the propagators (21) and (26) and the normalization constants (22) and (27) it turns out that the couplig constant $\alpha$ and the mass $m$ always enter in the combination $m\alpha$ for pseudoscalar and $\sqrt{m}\alpha$ for scalar fields. So it might be possible that this entity is the same for all mesons. This would relate the coupling constant $\alpha_\pi$ of the pion to the coupling constants of the other mesons setting:

$$\begin{aligned} \alpha(m) &= \alpha_\pi \cdot \left(\frac{m_\pi}{m}\right)^{\frac{1}{2}} \qquad \text{for scalar fields} \\ \alpha(m) &= \alpha_\pi \cdot \left(\frac{m_\pi}{m}\right) \qquad \text{for pseudoscalar and vector fields.} \end{aligned} \qquad (28)$$

These functions $\alpha(m)$ compared with the values of $\alpha$ for the different mesons obtained by fitting the modified solitary wave propagator to the propagators and form factors of the Bonn–B parameter set are plotted in figure [5]. The agreement with the empirical scaling law (28) is as impressive as unexpected.

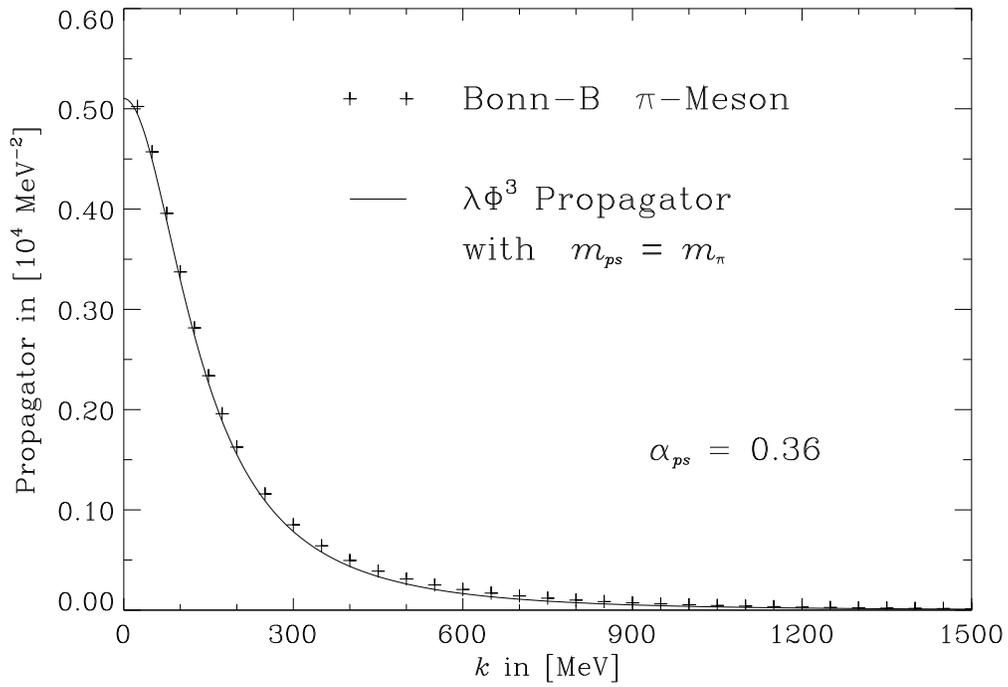

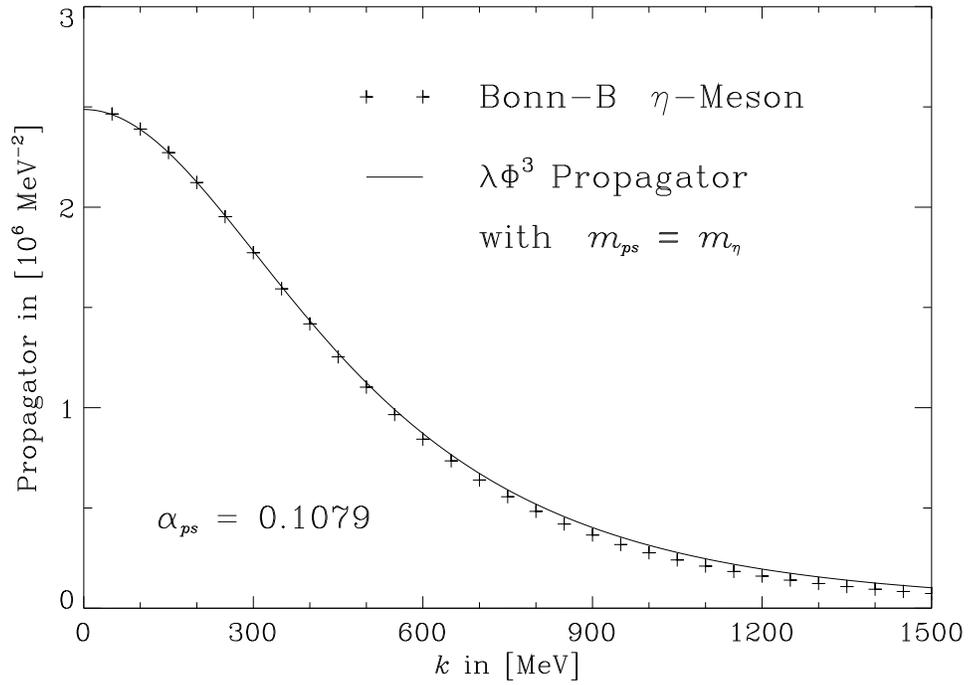

Figure 2: Comparison of pseudoscalar Meson Bonn–B and $\lambda\Phi^3$ Propagators

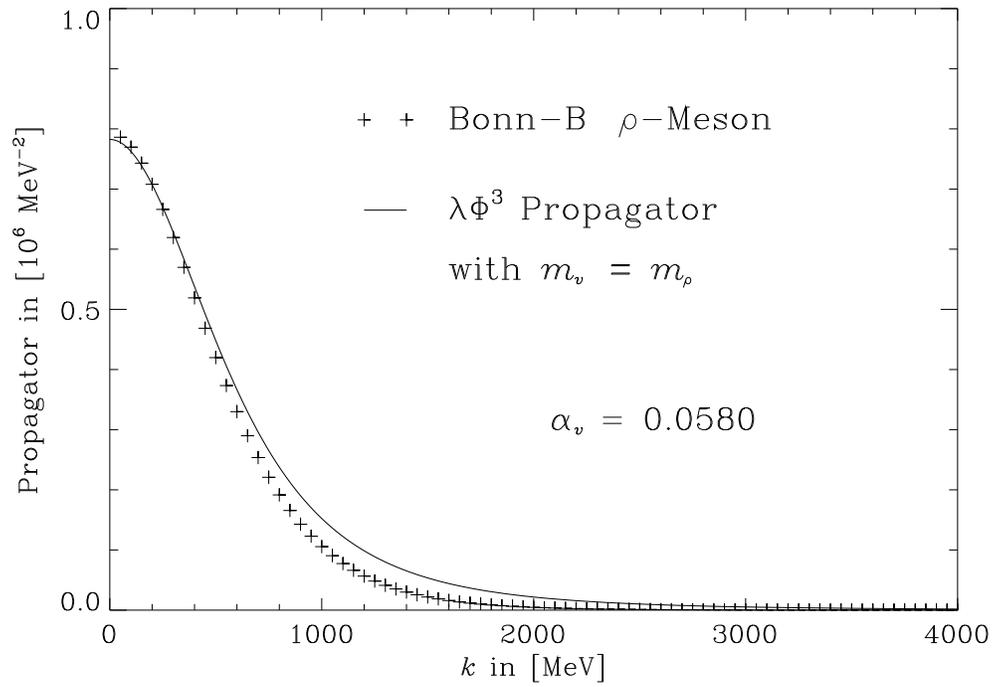

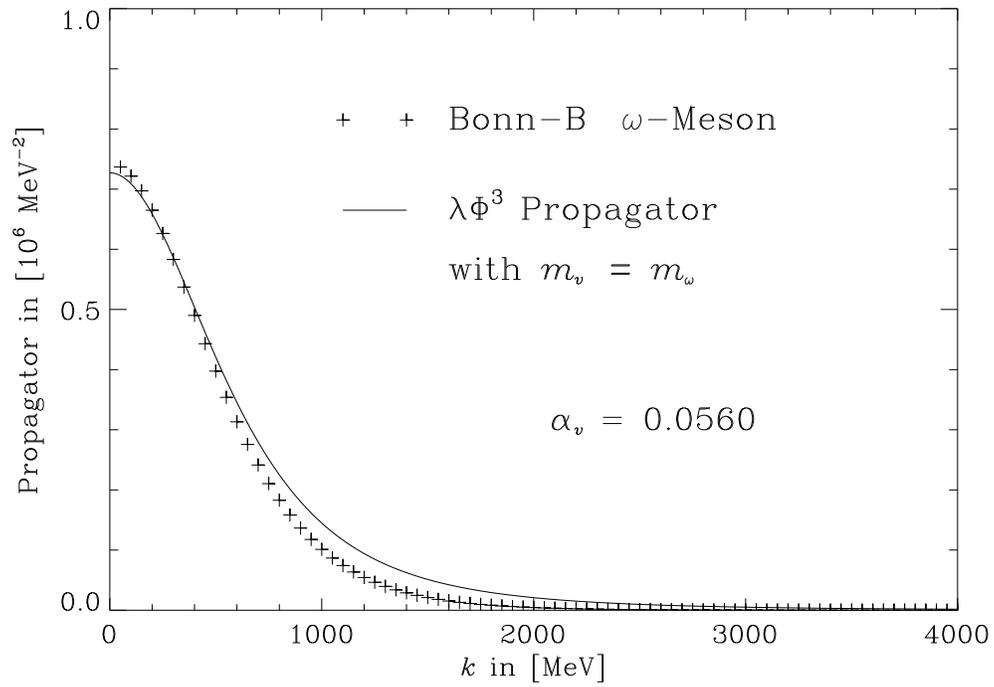

Figure 3: Comparison of vector Meson Bonn–B and $\lambda\Phi^3$ Propagators

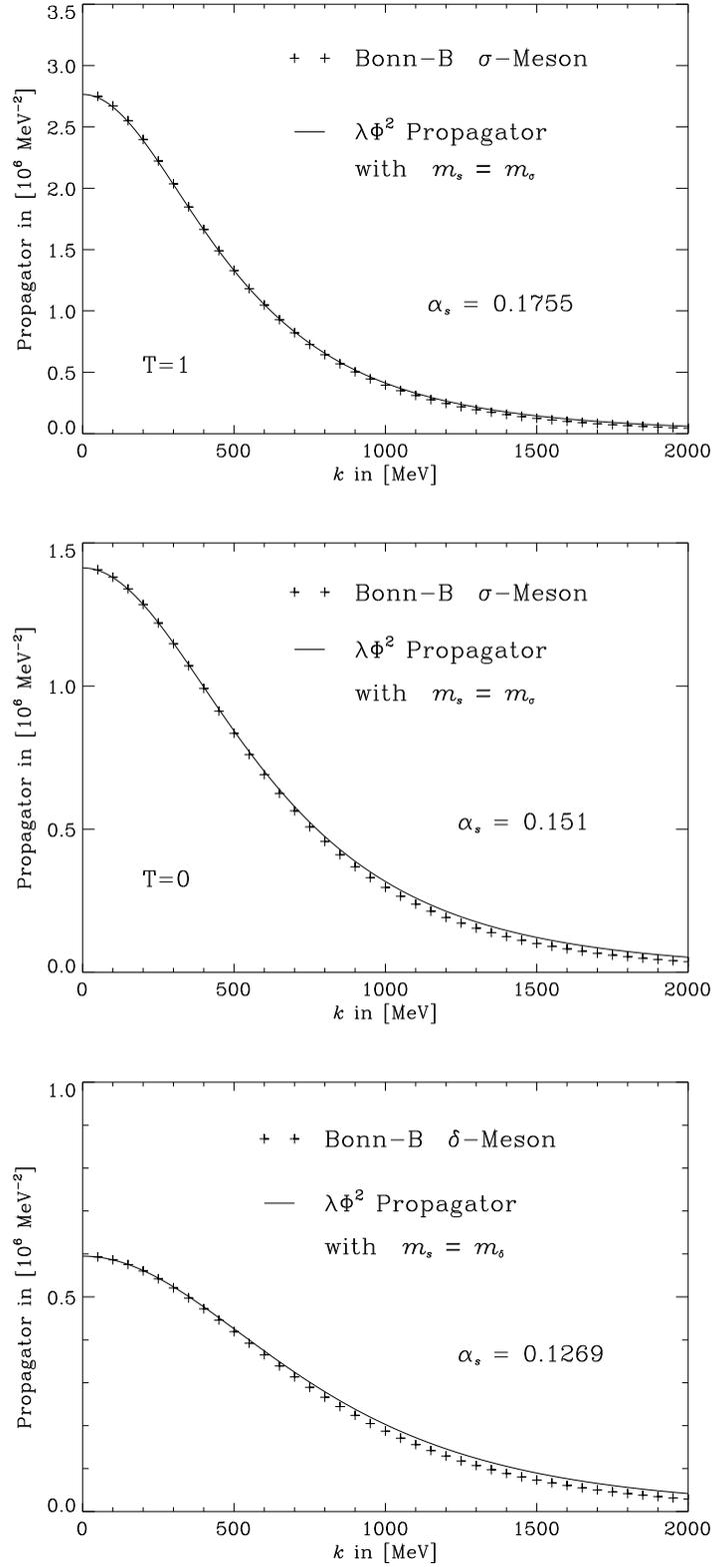

Figure 4: Comparison of scalar Meson Bonn–B and $\lambda\Phi^2$ Propagators

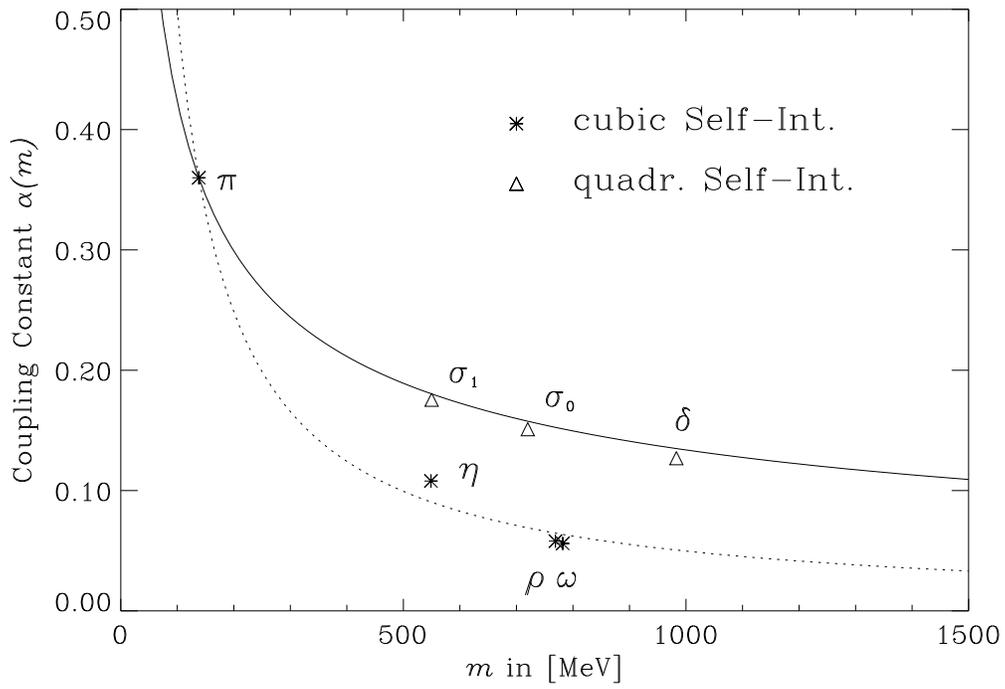

Figure 5: An empirical scaling law for the coupling constant depending on $m$

# 4 Conclusions

The model of solitary waves is able to describe self–interacting meson fields in a consistent way. Proper normalization makes all self–scattering diagramms finite. When applied in NN–scattering the propagator fits the Bonn–B propagator multiplied with the form factor. We therefore conclude that in a model using solitary waves *form factors are not necessary*. In addition there seems to be a connection between the coupling constants in form of a scaling law (28). This would allow to reduce all parameters to the pion coupling constant. Instead of the seven cut–off masses and the seven exponents $n_\beta$ this would yield a consistent description of all mesons with one fundamental coupling constant.

To make a crude test of the scaling law (28) one can calculate the meson coupling constants from the pion coupling constant and compare the obtained propagators with the Bonn–B propagators. The cut–off masses can then be adjusted to fit the solitary wave propagator. So one gets a set of seven cut–off parameters which correspond to one coupling constant $\alpha_\pi$. Using the value:

$$\alpha_\pi = 0.36$$

yields:

|                      | $\pi$ | $\eta$ | $\rho$ | $\omega$ | $\sigma_1$ | $\sigma_0$ | $\delta$ |
|----------------------|-------|--------|--------|----------|------------|------------|----------|
| $\Lambda_\beta$(GeV) | 1.7   | 1.3    | 1.95   | 1.95     | 1.9        | 1.95       | 2.1      |

If one inserts these parameters into the original Bonn–B potential and calculates the low energy phase shifts one gets qualitative agreement with e.g. the Arndt SM94 phase shift analysis (fig. [6]). This is a promising result and from this the proper normalized solitary wave exchange potential PSWEP deserves further investigations since one can hope to obtain a consistent qualitative description of NN data based on very few physically motivated parameters.

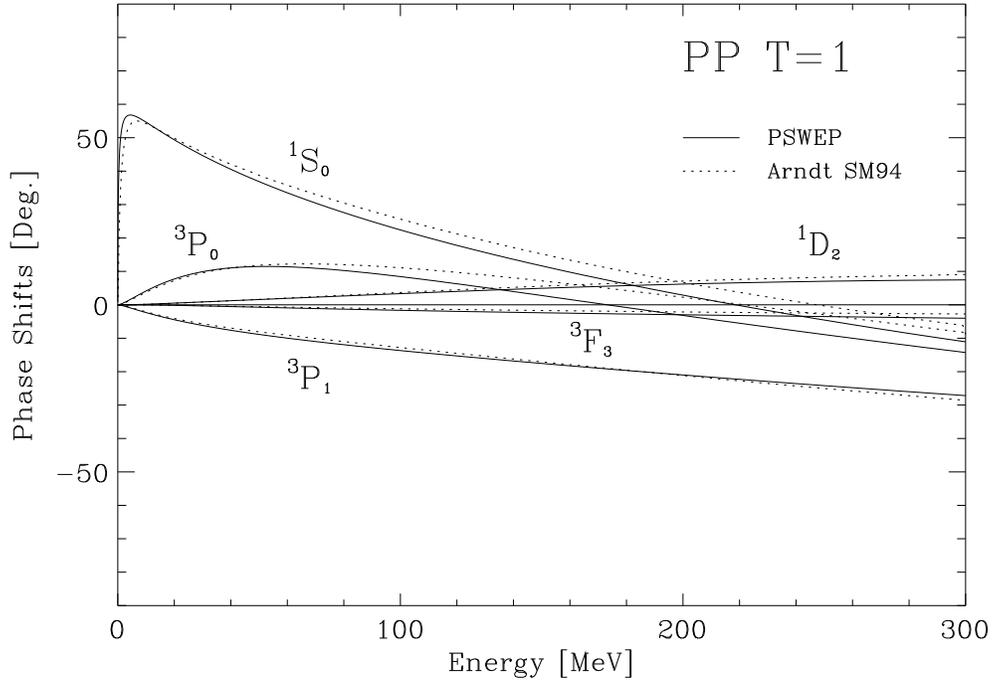

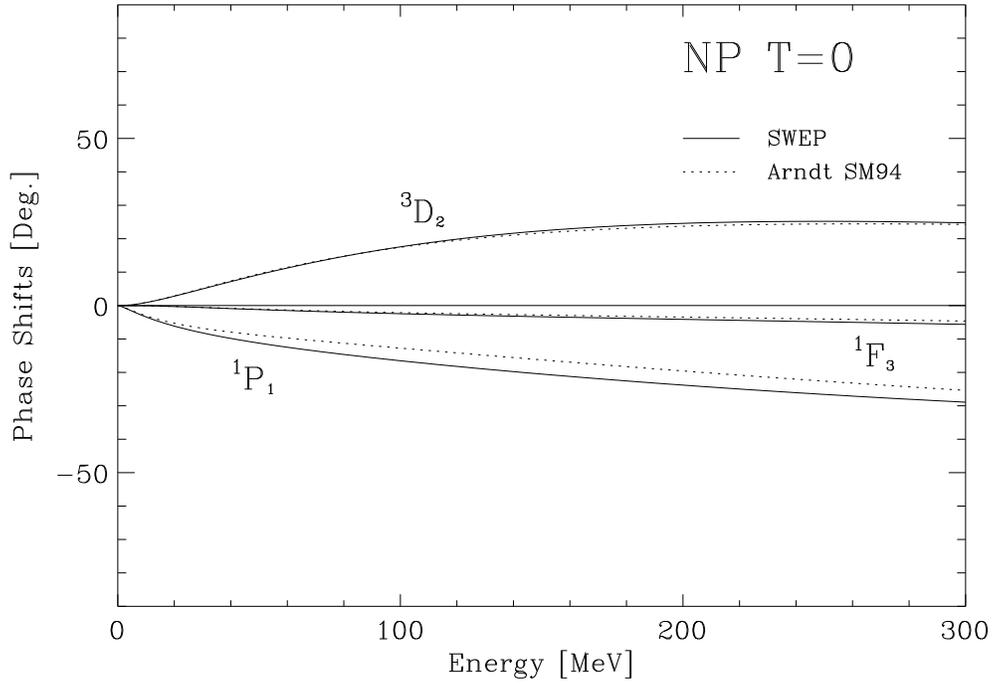

Figure 6: Phase shifts obtained from the Bonn OBEP using the parameter set predicted by $\alpha_\pi = 0.36$